\documentclass[]{nejsds}

\volume{0}
\issue{0}
\pubyear{2024}
\articletype{research-article}
\doi{10.51387/24-NEJSDS69}
\firstpage{1}
\lastpage{16}

\usepackage{amssymb, amsmath, amsthm}
\usepackage{bm}
\usepackage{booktabs, threeparttable, tabulary, caption, array}
\usepackage{algorithmicx}
\usepackage{algorithm}
\usepackage{algpseudocode}

\newcommand*\diff{\mathop{}\!\mathrm{d}}

\algnewcommand{\LeftComment}[1]{\Statex \(\triangleright\) #1}

\theoremstyle{remark}

\graphicspath{{./}{./plots/}}
\allowdisplaybreaks

\usepackage[]{lineno}
\newcommand*\patchAmsMathEnvironmentForLineno[1]{%
	\expandafter\let\csname old#1\expandafter\endcsname\csname 
	#1\endcsname
	\expandafter\let\csname oldend#1\expandafter\endcsname\csname 
	end#1\endcsname
	\renewenvironment{#1}%
	{\linenomath\csname old#1\endcsname}%
	{\csname oldend#1\endcsname\endlinenomath}}%
\newcommand*\patchBothAmsMathEnvironmentsForLineno[1]{%
	\patchAmsMathEnvironmentForLineno{#1}%
	\patchAmsMathEnvironmentForLineno{#1*}}%
\AtBeginDocument{%
	\patchBothAmsMathEnvironmentsForLineno{equation}%
	\patchBothAmsMathEnvironmentsForLineno{align}%
	\patchBothAmsMathEnvironmentsForLineno{flalign}%
	\patchBothAmsMathEnvironmentsForLineno{alignat}%
	\patchBothAmsMathEnvironmentsForLineno{gather}%
	\patchBothAmsMathEnvironmentsForLineno{multline}%
      }

\begin{document}
\begin{frontmatter}
\pretitle{Research Article}

\title{Heteroscedastic Growth Curve Modeling with Shape-Restricted Splines}

\begin{aug}
\author[a]{\inits{J.}\fnms{Jieying} \snm{Jiao}\ead[label=e1]{jieying.jiao@uconn.edu}}
\author[b]{\inits{W.}\fnms{Wenling}
  \snm{Song}\ead[label=e2]{songwenlingcarol@163.com}}
\author[c]{\inits{Y.}\fnms{Yishu} \snm{Xue}\ead[label=e3]{yishu.xue@uconn.edu}}
\author[d]{\inits{J.}\fnms{Jun} \snm{Yan}\thanksref{c1}\ead[label=e4]{jun.yan@uconn.edu}}
\thankstext[type=corresp,id=c1]{Corresponding author.}
\address[a]{Department of Statistics, \institution{University of
    Connecticut}, \cny{USA}.\\ \printead{e1}}
\address[b]{Department of Obstetrics, \institution{The First Hospital of Jilin University}, \cny{China}.\\ \printead{e2}}
\address[c]{Department of Statistics, \institution{University of
    Connecticut}, \cny{USA}.\\ \printead{e3}}
\address[d]{Department of Statistics, \institution{University of
    Connecticut}, \cny{USA}.\\ \printead{e4}}
\end{aug}

%\dedicated{Dedicated to}

\begin{abstract}
Growth curve analysis (GCA) has a wide range of
  applications in various fields
  where growth trajectories need to be modeled. Heteroscedasticity is
  often present in the error term, which can not be handled with
  sufficient flexibility by standard linear fixed or mixed-effects
  models. One situation that has been addressed is where the error
  variance is characterized by a linear predictor with certain
  covariates. A frequently encountered
  scenario in GCA, however, is one in which the variance
  is a smooth function of the mean with known shape restrictions.
  A naive application of
  standard linear mixed-effects models would underestimate the
  variance of the fixed effects estimators and, consequently,
  the uncertainty of the estimated growth curve. We propose to model
  the variance of the response variable as a shape-restricted
  (increasing/decreasing; convex/concave)
  function of the marginal or conditional mean using shape-restricted splines.
  A simple iteratively reweighted fitting algorithm that takes advantage
  of existing software for linear mixed-effects models is developed.
  For inference, a parametric bootstrap procedure is recommended.
  Our simulation study
  shows that the proposed method gives satisfactory inference with
  moderate sample sizes. The utility of the method is demonstrated
  using two real-world applications.
\end{abstract}


\received{\smonth{5} \syear{2023}}
\end{frontmatter}

\section{Introduction}\label{sec:Intro}

Growth curve analysis (GCA) plays a critical role in various fields
such as agronomy~\citep{strydhorst2018plant}, animal
science~\citep{robert2002modelling}, 
biology~\citep{sandland1979stochastic},
clinical trials~\citep{zee1998growth}, and psychological
studies~\citep{Curran:twelve:2010, mcardle2003growth,
  carter1992random}, among others. A GCA provides information about
not only the mean but also the variation of the growth trend of a
certain population. For example, reference growth charts for
children's height, weight, and other physical characteristics are
widely used in wellness checks. A growth chart typically depicts a
collection of quantiles of the distribution of physical
characteristics of the reference population as a function of age.
Accurate characterizations of the growth trajectory
in both the mean level and the variation level
are needed to make valid inferences and draw meaningful conclusions.
The mean level of a growth trajectory has been extensively studied
with a variety of functional forms such as fractional
polynomial~\citep{healy1988distribution, royston1998method} and
smoothing splines~\citep{cole1992smoothing}. In contrast, the
variation has been studied but far less
extensively.

Heteroscedasticity is a commonly encountered challenge in GCA. We
often observe larger variance as the mean gets bigger or as the growth
pattern proceed with time.
The error variance can be modeled as a smooth function of time or the
mean response. Kernel-based methods have been used, which led to
uniformly consistent estimator of the variance
function~\citep{carroll1982adapting, muller1987estimation}.
Covariates could be incorporated into the variance by an
additional regression for the dispersion \citep{smyth1989generalized,
  nelder1998joint}. For predictive purposes, a parametric distribution
at any time point is often desired. The lambda-mu-sigma (LMS) method
handles heteroscedasticity along with non-normality. In particular, it
assumes that, after being standardized by a time-specific median~$\mu$
and Box--Cox transformed with a time-specific power~$\lambda$, the
response follows a normal distribution with mean zero and
time-specific standard deviation~$\sigma$~\citep{cole1988fitting,
  cole1992smoothing}. The functions $\mu$, $\sigma$, and $\lambda$ are
assumed to evolve smoothly with time, which can be modeled by splines
of time. Distributions other than the normal distribution can be used
with time-specific parameters in the generalized additive modeling
framework for location, scale, and shape (GAMLSS)~\citep{rigby2005generalized,
  rigby2013discussion, rigby2014automatic}. Quantile regression is a
distribution-free method that directly models
the age-specific quantiles of the response, possibly
conditioning on covariates~\citep{wei2006quantile}.

Clustered data, which often arise in GCA, bring an additional
challenge of handling the within-cluster dependence. The number of
repeated measures over time on the same subject can be as
many as 30--40~\citep{Stone:etal:science:2007}.
The linear mixed-effects model (LMM) introduces within-cluster
dependence through cluster-level random effects. Splines can be used
to get a non-parametric estimation of the growing
curve~\citep{rice2001nonparametric, huang2002varying,
    huang2004polynomial, wu2018nonparametric}. An additional model on the
variance or scale leads to the
mixed-effects location-scale model~\citep{Robert:etal:modelling:2002,
  Hedeker:etal:application:2008, Hedeker:etal:modeling:2012}.
The variance model is formulated with covariates or through a
scale-mixture with, for example, an inverse gamma scale.
The \texttt{lme()} function in \textsf{R} package
\textbf{nmle}~\citep{Pinheiro:nlme:2022} can
fit LMMs with heteroscedasticity in a set of
pre-programmed forms~\citep[p.71--100]{Zuur:etal:Mixed:2009}, and the
best form can be selected using Akaike information
criterion (AIC)~\citep{akaike1998information} or Bayesian information criterion
(BIC)~\citep{schwarz1978estimating}. The covariance structure can be modeled
directly with nonparametric methods for flexible
shapes~\citep[e.g.,][]{diggle1998nonparametric}.
The normal distribution assumption of the response variable can be
relaxed by using GAMLSS with random effects in the additive
terms~\citep[p.247--252]{stasinopoulos2017flexible}. The generalized
estimating equation (GEE) method~\citep{liang1986longitudinal} focuses
on the marginal modeling. It has been generalized to handle
heteroscedasticity and within-cluster
correlations~\citep{Yan:Fine:esti:2004}, but marginal models in
general are not suitable for subject-specific predictions.

Despite the extensive GCA literature, there are two limitations in routine
analyses. The first is that there is no convenient way to put shape
restrictions, such as monotonicity and/or convexity/concavity, on
the variance in addition to the mean of a growth curve.
Existing methods such as GAMLSS~\citep{stasinopoulos2017flexible}
allow flexible shapes in the mean or variance
structure, but there is no direct way yet to ensure shape restrictions. If
effectively used, such restrictions could improve the efficiency in inferences.
The second limitation is that the existing methods do not have the flexibility
to allow the variance to be modeled as a function of the mean, which is common
in generalized linear models. The R package \textbf{nmle}
\citep{Pinheiro:nlme:2022} only allows the
mean for independent data and the marginal mean for clustered data in
the variance structure, but not the conditional
mean given random effects for clustered data. Variance as a function of the mean
could, again, improve the efficiency in inferences, especially when the mean
depends on multiple covariates.

To break the aforementioned two limitations, we propose to model the
variance in GCA as a shape-restricted function of the growth
level~\citep{lambert2001analysis}. The shape restrictions include
monotonicity and/or convexity/concavity, accommodated with
shape-restricted splines such as monotone splines~\citep{Ramsay:monotone:1988}
or convex splines~\citep{Meyer:inference:2008} with evenly spaced
knots and constrained parameters. This can enable us to tickle the
scenario where variance gets bigger at the end or beginning of the
growth curve. For clustered data, the variance
model can incorporate the growth level either through the marginal
mean or the conditional mean 
given the cluster-level random effects. Either AIC or BIC can be used
to select the degrees of freedom of the splines and to select between
marginal and conditional mean models. The parameters are estimated in
an iteratively reweighted fitting algorithm. The performance of the
proposed methods is validated through an extensive simulation study
and applications to two real examples.

The rest of this paper is organized as follows. 
Section~\ref{sec:ISpline} gives a review of
the shape-restricted spline basis. Growth models with shape-restricted
Heteroscedasticity for both independent and clustered data are
presented in Section~\ref{sec:Methods}. A simulation study is reported
in Section~\ref{sec:simu} to assess the performance of the methods. We
illustrate the use of the proposed approach with the fetal pancreas
length data and the chicken weight data in
Section~\ref{sec:real}. A discussion concludes in
Section~\ref{sec:discussion}. The computing code is publicly available at
\url{https://github.com/JieyingJiao/GCA_Code}.

\section{Shape-Restricted Splines}\label{sec:ISpline}

Splines are piecewise polynomials, differentiable up to a certain
degree. They offer great flexibility in approximating unknown smooth
curves, and is often preferred to simple polynomial basis. It can give
similar results to polynomial basis even with a lower degree, while
avoiding the Runge's phenomenon for higher degree.

Applying splines to independent or clustered data such as longitudinal
data has been extensively studied in the
literature, such as B-spline~\citep{rice2001nonparametric, huang2002varying,
  huang2004polynomial, wu2018nonparametric}. There are other type of
splines that have certain shape restrictions, such as monotonicity and
convexity. Using shape-restriction splines to estimate smooth curves
with certain shapes hasn't been discussed before.

Specifically, a shape-restricted curve is approximated
by a linear combination of a set of shape-restricted spline bases,
where the coefficients are restricted to get the desired pattern. Before
introducing our proposed method, which employs the I-spline
bases and C-spline bases, we first briefly review how they are
constructed.

To define shape-restricted spline bases, we start from M-splines.
M-spline bases are standardized versions of B-spline bases so that
they integrate to 1~\citep{curry1966polya}. An M-spline of degree~$k$
over an interval $[l, u]$ is defined recursively as
\begin{align*}
    M_i^{(1)}(x) &=
    \begin{cases}
      \displaystyle \frac{1}{t_{i+1} - t_i}, & t_i \le x \le t_{i+1},\\
      0, &\mbox{otherwise},
    \end{cases}\\
    M^{(k)}_{i}(x) &=
    \begin{cases}
      \displaystyle \frac{k[(x-t_i)M^{(k-1)}_i(x) +
        (t_{i+k} - x)M^{(k-1)}_{i+1}(x)]}
      {(k-1)(t_{i+k} - t_i)},
      & t_i \le x \le t_{i+k}\\
      0, &\mbox{otherwise},
    \end{cases}
\end{align*}
$i = 1, \ldots, m + 2k$, where $t_i$'s are the knots with
\[
  l = t_1 = \dots = t_k < \dots <
  t_{m+k+1} = \dots = t_{m+2k} = u,
\]
and $m$ is the number of internal knots.
The M-spline bases are positive over~$[l, u]$. A linear
combination of M-spline bases with nonnegative coefficients is
non-negative. Same as B-splines, it is continuously differentiable
up to $k - 1$ times for $k \ge 1$.

I-splines are integrals of M-splines~\citep{Ramsay:monotone:1988}. The
I-spline bases with degree~$k$ over the interval~$[l, u]$ are
\begin{equation*}
  I^{(k)}_i(x) =
  \int_{l}^x M^{(k)}_i(s) \diff s,
  \quad
  l \le x \le u,
  \qquad
  i  = 1, \ldots, m + 2k. 
\end{equation*}
Because their derivatives are M-splines, which are non-negative,
I-spline bases can be used for modeling monotonic functions. A linear
combination of I-spline bases with non-negative (or non-positive) coefficients
are non-decreasing (or non-increasing). An intercept is always needed
when using I-spline bases since their lowest order is linear.

C-splines are integrals of I-splines~\citep{Meyer:inference:2008}. The
C-spline bases with degree~$k$ over the interval~$[l, u]$ are
\begin{equation*}
  C^{(k)}_i(x) =
  \int_{l}^x I^{(k)}_i (s) \diff s,
  \quad
  l \le x \le u,
  \qquad
  i = 1, \ldots, m + 2k.
\end{equation*}
This set of bases does not have a linear or a constant term, both of
which need to be added when fitting curves. With restrictions on the
coefficients, C-splines can be used to approximate functions
with specific combinations of monotonicity (increasing or decreasing)
and shape
(convexity or concavity)~\citep{wang2021shape}. A commonly seen pattern 
in growth curve is non-decreasing concave, which can be implemented by
restricting the first derivatives to be positive and second derivatives to be
negative.

In implementation, we used the I-spline and C-spline bases
from \textsf{R} package \textbf{splines2}~\citep{wang2021shape}.
As illustrated later, the degrees of freedom can be chosen by AIC or BIC. For a
typical GCA, a moderately complicated pattern can be approximated by spline
bases with a few (3--5) degrees of freedom.

\section{GCA with Shaped-Restricted
  Heteroscedasticity}\label{sec:Methods}
Although clustered or longitudinal data is often encountered for GCA,
there are also situations that only one measure is collected from each
subject, such as the pancreas data presented in
Section~\ref{sec:real}. The shape restrictions in mean and error
terms can exist in both data types in GCA, but haven't been
systematically discussed.

The proposed method can be applied to either linear regression model
for independent data or LMM for clutered data, depending on if there are
repeated measurements on same subject. For clarity of presentation, we
start from the independent data setting which is simpler, and
then consider the more complicated clustered data setting which is
also more common in GCA. Inferences and model selection come next.

\subsection{Model for Independent Data}

Suppose the data is collected from $n$ subjects, and each of the
subjects was only observed once at a random time point. Specifically,
let the observed data or measurement for the $i$th subject be $y_i$,
and the observed time be $t_i$, $i = 1, 2, \dots, n$. Since $y_i$ are
from different subjects, they are independent to each
other. Additional information except time are possible, such as gender
or treatment group. They are represented by a $p$-dimensional
$\textbf{x}_i$ for the $i$th subject. To introduce heteroscedasticity, we use a
smooth function~$g(\nu_i)$ to characterize the standard deviation of the
regression error of the $i$th subject, where~$v_i$ is some index
variable. The index variable can be observed such as time, or unobserved such
as the mean of the corresponding subject from the linear model. The smooth
function~$g(\cdot)$ areis parametrized by shape-restricted splines.

The growth pattern against time is often non-linear and present
certain shape restrictions, such as increasing with time. This can be
realized by using shape-restricted spline of time in the mean
model. Once the degree and degree of freedom of spline bases being
chosen, it can be represented using the same format as the parametric
part: linear combination of coefficient and spline bases. Using spline
bases in the mean pattern has been discussed extensively in the
previous work~\cite{rice2001nonparametric, huang2002varying,
  huang2004polynomial, wu2018nonparametric}, and our focus is more on
using the spline on the variance part. For the
simplicity of expression, we choose to include the spline bases of
time as part of the vector $\textbf{x}_i$ instead of explicitly show
them separately. The process to chose the degree and degree of freedom
is the same as for the splines in the error term. 

Using I-splines as an example, a heteroscedastic linear model is
\begin{equation}
  \begin{split}
    y_i &= \textbf{x}_i^\top \bm{\beta}+\varepsilon_i, \quad
    i=1,\ldots, n,\\
    \varepsilon_i &\sim \mathrm{N}(0, g^2(v_i, \bm{\theta})),\\
    g(v_i;\bm{\theta}) &= \theta_0 + \sum_{k = 1}^{K} \theta_{k}
    I_{2, k}(v_i),
  \end{split}
  \label{ind_model}
\end{equation}
%$i = 1, \ldots, n$, 
where~$\bf{\beta}$ is a $p$-dimensional regression
coefficient vector for~$\textbf{x}_i$,
$\varepsilon_i$ is the normally distributed regression error with mean zero and
standard deviation $g(\nu_i; \bf{\theta})$, $\{I_{1,k}(\cdot), k =
1, \ldots, K\}$ are I-spline bases with $K$ degrees of
freedom, and ${\bf\theta} =
(\theta_0, \dots, \theta_{K})$ is a $(K+1)$-dimensional
coefficient vector. The degree and degrees of freedom for each spline
bases need to be selected using model selection method introduced in
Section~\ref{sec:inference}, and the internal knots are evenly
spaced. The coefficients ${\bf\theta}$ can be restricted to control the shape of the
heteroscedasticity as a function of $\nu_i$. For example, if the variance
increases with the mean (or time), the coefficients ${\bf \theta}$ can be
restricted to be non-negative.

If concavity or convexity is desired, the I-splines can be replaced with
C-splines and a linear term of time with appropriate restrictions on the
coefficients, as introduced in Section~\ref{sec:ISpline}. Interaction
terms can be introduced in the mean function to allow the covariates
have time-varying coefficients~\citep{huang2002varying,
  huang2004polynomial}.

\subsection{Model for Clustered  Data}

When more than one measures were collected from each of the~$n$ subjects, the
observed data will have a clustered structure. Let the number of
repeated measures on the $i$th subject be $n_i$, and it might be
different for each subject and sometimes might be small or even just
1. The $j$th observation of the $i$th subject
is $y_{i,j}$ which is collected at time $t_{i, j}$, where $j = 1,
\dots, n_i$, $i = 1, \dots, n$. Same as before, we still use
spline bases to estimate the error variance in order
to put the shape restrictions, but with a linear mixed-effects
model to account for the dependence structure within the dataset.
We use the matrix notation for simplicity
of demonstration:
\begin{equation*}
    {\bf y}_i = \begin{pmatrix}
      y_{i,1}\\
      .\\
      .\\
      .\\
      y_{i,n_i}\\
    \end{pmatrix},\quad
    {\bf X}_i = \begin{pmatrix}
      \textbf{x}_{i,1}^\top\\
      .\\
      .\\
      .\\
      \textbf{x}_{i, n_i}^\top\\
    \end{pmatrix}, \quad
    \bm{\varepsilon}_{i} =
    \begin{pmatrix}
      \varepsilon_{i, 1}\\
      .\\
      .\\
      .\\
      \varepsilon_{i, n_i}
    \end{pmatrix}.
\end{equation*}
Again using I-Splines as an example, the model for the~$i$th subject is
\begin{equation}
  \begin{split}
    {\bf y}_i
    &=
    {\bf X}_i\bm{\beta} + \mathbf{Z}_i{\bf b}_i+\bm{\varepsilon}_{i}
    , \quad i=1,\ldots, n,\\
    % {\bf\varepsilon}_i' &= (\varepsilon_{i1}, \varepsilon_{i2}, \dots,
    % \varepsilon_{in_i}),\\
    \bm{\varepsilon}_{i} &\sim \mathrm{MVN}\left(\bm{0},~
\mathrm{diag}(g^2(v_{i,1}, \bm{\theta}), \dots, g^2(v_{i,n_i}, \bm{
\theta}))\right),\\
    g(v_{i,j}, \bm{\theta}) &= \theta_0 + \sum_{k=1}^{K}\theta_{k}
    I_{k}(v_{i,j}), \quad j = 1, \ldots, n_i,
  \end{split}
  \label{clustered_model}
\end{equation}
where $\textbf{x}_{i,j}$ is a~$p$ dimensional covariate
vector for fixed effects which can include the spline bases of time,
$\mathbf{Z}_i$ is an~$n_i \times q$ design matrix for random effects,
$\mathbf{\beta}$ is a~$p$ dimensional fixed effects vector,
$\mathbf{b}_i$ is a~$q$-dimensional random effects vector with
covariance matrix~$\mathbf{B}$ parameterized by vector $\bm{\alpha}$,
and MVN is the multivariate normal distribution. Shape restrictions
can be applied on the coefficient of spline bases, and $K$ need to be selected
using the model section method in Section~\ref{sec:inference}. Other
notations are the clustered analogs to those in
Equation~\eqref{ind_model}.

A special choice for the index variable $v_{ij}$ is the mean of the
response variable. For a mixed-effects model, this mean can be
conditional on the random effects or not. If random effects are conditioned on,
the mean is
\begin{equation}
\bm{\mu}_{i,\mbox{c}} = E[{\bf y}_i\mid {\bf b}_i] = {\bf
  X}_i\bm{\beta} +
    {\bf Z}_i{\bf b}_i;
  \label{eq:cmean}
\end{equation}
otherwise it is
\begin{equation}
    \bm{\mu}_{i,\mbox{m}} = E[{\bf y}_i] = {\bf X}_i\bm{\beta}.
  \label{eq:mmean}
\end{equation}
For ease of referencing, we call them conditional mean and marginal mean,
respectively.
When the error variance changes with the marginal mean, the response variable
still has a multivariate normal distribution. If the conditional mean is in the
error variance structure, there is dependence between the random effects and
the
error term, and the response variable no longer has a multivariate normal
distribution. To calculate the likelihood function for this situation, as
needed
in AIC and BIC calculations, numerical integration is needed. See details
in Section~\ref{sec:inference}.

Same as for the indepdent data, C-splines can be used for concavity or
convexity shape restriction, and interaction terms in the mean
function can allow time-varying coefficients for the covariates.

\subsection{Inference}
\label{sec:inference}

\begin{algorithm}[tbp]
  \caption{Iteratively reweighted fitting algorithm for clustered data.}
\begin{algorithmic}[1]
  \LeftComment {\textbf{Input} $\{\mathbf{y}_i, \mathbf{X}_i,
    \mathbf{t}_i, \mathbf{Z}_i, i =
1, \dots, n\}$.}
  \Procedure{}{}
  \State Fit a linear mixed-effects model without weight.
  \State Get estimate $\hat{\bm{\beta}}$ of $\bm{\beta}$, 
  estimate $\hat{\bm{\alpha}}$ of $\bm{\alpha}$,
  residuals $\bm{e}_i$, and fitted (marginal or conditional) mean
  $\hat{\bm{\mu}}_i$ $i = 1, 2, \dots, n$.
  \Repeat
  \State Treat residuals $\bm{e}_i$'s as an observation from
  $\mathrm{N}(0,  g^2(\hat{\bm{\mu}}_i, \bm{\theta}))$, $i = 1, 2, \dots, n$.
  \State Get maximum likelihood estimate $\hat{\bm{\theta}}$ 
  with monotone constraints that $\bm{\theta} > 0$.
  \State Fit a linear mixed-effects model with weight
  $\{g^{-1}(\hat{\bm{\mu}}_i,  \hat{\bm{\theta}}),\ i = 1, 2, \dots, n\}$
  \State Get updated $\hat{\bm{\beta}}$, $\hat{\bm{\alpha}}$, $\bm{e}_i$,
  and $\hat{\bm{\mu}}_i$, $i = 1, 2, \dots, n$.
  \Until{$\hat{\bm{\beta}}$ converges}.
    
  \EndProcedure
  \LeftComment {\textbf{Output}
$\hat{\bm{\beta}}$, $\hat{\bm{\alpha}}$, and $\hat{\bm{\theta}}$.}
\end{algorithmic}
\label{algorithm_estimation}
\end{algorithm}

The maximum likelihood method can be used to get parameter estimates in theory
as long as appropriate restrictions on the coefficients are imposed to enforce
the shape restrictions. To obtain the maximum likelihood estimator, we propose
an iteratively reweighted fitting procedure that takes advantage of existing
software packages for linear mixed-effects models allowing
weights. This method is flexibile to deal with different scenatios
including the error variance changing with conditional or marginal
mean. It can also be easily computed since no closed-form solutions
need to be derived. The steps
are summarized in Algorithm~\ref{algorithm_estimation} for clustered data when 
the error variance is changing with the mean. We use~$\bm{\mu}_i$ in the
algorithm to represent either the conditional mean or the marginal mean,
and~$\hat{\bm{\mu}}_i$ for the estimated value of the mean. Algorithm
for independent data is similar and simpler, and will not be repeated
here. The shape restrictions on the heteroscedasticity (and the
mean model) can be imposed with a constrained optimizer, such as the
\texttt{constrOptim()} function in R.

To construct reference quantiles in a GCA, we suggest using
parametric bootstrap. This is very similar to the resampling-subject
bootstrap (RSB) method~\citep{huang2002varying, wu2018nonparametric}
since the bootstrap sample is generated on subject level to maintain
the cluster structure. The main difference is that our bootstrap
sample is generated with estimated parameter, instead of the
residuals. This is because the error variance in our model is
estimated with spline bases.
The bootstrap method avoids deriving the likelihood function and the
Hessian matrix ~\citep[][p.~227]{rossi2018mathematical}, which is
challenging when the error variance changes with the conditional mean. This is
of particular importance when some of the estimated $\theta$'s are on
the boundaries of the constrained parameter space~\citep{Andrews:est:1999}.
Since we focus on the final fitted curve instead of the basis coefficients,
bootstrap provides a natural solution for the quantitles of the fitted curve
regardless of whether some of the estimated $\theta$'s are on the boundaries.

\begin{algorithm}[tbp]
  \caption{Steps to get one parametric bootstrap sample for clustered data.}
  \begin{algorithmic}[1]
    \LeftComment {\textbf{Input} $\hat{\bm{\beta}}$, 
  $\hat{\bm{\alpha}}$, and $\hat{\bm{\theta}}$.}
  \Procedure{}{}
  \State Generate random effects
  $\bm{b}^*_i$'s from $\mathrm{N}(0, \mathbf{B}(\hat{\bm{\alpha}}))$,
  $i = 1, 2, \dots, n$.
  \State Let $\bm{\mu}^*_{i} = (\mu^*_{i1}, \dots, \mu^*_{in_i})$ be
  ${\bf X}_i\hat{\bm{\beta}}+ {\bf Z}_i{\bf b}^*_i$
  when use conditional mean, or
  ${\bf X}_i\hat{{\bf\beta}}$ when use marginal mean.
  \State Generate error terms ${\bm\varepsilon}_{i}^*$ from
$\mathrm{MVN}\left(\bm{0},~\mathrm{diag}(g^2(\mu^*_{i1}, \hat{\bm{\theta}}),
\dots,
    g^2(\mu^*_{in_i}, \hat{\bm{\theta}}))\right)$, $i = 1, 2, \dots, n$.
  \State Let $\bm{y}_i^* = {\bf X}_i\hat{\bm{\beta}}+
    {\bf Z}_i{\bf b}^*_i + \bm{\varepsilon}_i^*$, $i = 1, 2, \dots, n$.
  \State Apply Algorithm~\ref{algorithm_estimation}
  to $\{\bm{y}_i^*, \mathbf{X}_i,\mathbf{t}_i, \mathbf{Z}_i, i = 1, 2, \dots, n\}$ and
  record the output $\hat{\bm{\beta}}^*$, $\hat{\bm{\alpha}}^*$, and
  $\hat{\bm{\theta}}^*$.
  \EndProcedure
  \LeftComment {\textbf{Output} One bootstrap copy
$\{\hat{\bm{\beta}}^*, \hat{\bm{\alpha}}^*, \hat{\bm{\theta}}^*\}$.}
\end{algorithmic}
\label{algorithm_inference}
\end{algorithm}

Algorithm~\ref{algorithm_inference} summarizes the steps to get one parametric
bootstrap sample for clustered data. Same as before, the algorithm for 
independent data will be similar and will not be displayed here. Repeating this
process gives a large sample of bootstrap copies of the point estimates of the
model parameters. Their empirical standard
deviations are then used as the standard errors of the model parameter
estimates.
For mixed-effects models, parametric
bootstrap method has better performance compared with bootstrap methods that
only re-sample observations or residuals, as it produces
more accurate standard deviation of estimated
parameters, and closer-to-nominal coverage rates for confidence
intervals~\citep{das1999some, wu2002study, thai2013comparison}.

After the model fitting process, model checking can be done using the
residuals. The standardized residuals, i.e., the residuals divided by the
estimated
error standard deviation, should follow a standard normal distribution, and
their normality can be checked visually using a normal Q-Q plot, or other
normality testing methods.

\subsection{Model Selection}
\label{sec:ModelSelection}
With the internal knots evenly spaced, the values of degree and degree of
freedom are needed to generate the spline bases. They should depend
on the sample size $n$ and the number of observations each
subject has $n_i$ for clustered data. Additionally, for clustered
data, candidate models can either have the
conditional mean or the marginal mean in the error variance
function in Model~\eqref{clustered_model}.

Different values of spline degree and degree of freedom, and the
choice of using conditional mean or marginal mean, will significantly
impact the model fitting results. The popular model selection criteria
for such problems are AIC and BIC~\citep{rice2001nonparametric} as
they consider both the fitting accuracy and the model
complexity. It can give similar results to the 'deleting
subject cross validation' method, and is faster
to compute~\citep{rice2001nonparametric,huang2002varying}. Details are
as follow:
\begin{equation*}
  \begin{split}
    \mbox{AIC} &=-2 \log L + 2P,\\
    \mbox{BIC} &= -2 \log L + P \log n,
  \end{split}
\end{equation*}
where~$L$ is the likelihood function of the fitted model, $P$ is the
number of parameters, and~$n$ is the sample size. Models
with smaller AIC or BIC are preferred.

The numerical integration is needed for the scenario when the
model has error variance changing with the conditional
mean~\eqref{eq:cmean}. From the definition in
Model~\eqref{clustered_model}, the error variance now
depends on the random effects, and the likelihood function should be: 
\begin{equation*}
  \begin{split}
  L &= 
\prod_{i = 1}^n\int_{-\infty}^{+\infty}f({\bf y}_i \mid {\bf b}_i) f({\bf b}_i)
\diff {\bf b}_i,\\
{\bf b}_i &\sim \mathrm{MVN} \left(\mathbf{0},~\hat{\mathbf{B}}\right),\\
{\bf y}_i \mid {\bf b}_i &\sim \mathrm{MVN}\left(\mathbf{X}_i\hat{\bm{\beta}}
+ \mathbf{Z}_i{\bf b}_i, ~g^2(\bf{v}_{i}, \hat{\bm{\theta}})\right),
\end{split}
\end{equation*}
where~$f({\bf y}_i \mid {\bf b}_i)$ is the probability density function (pdf)
of the response vector~${\bf y}_i$ conditioning on the random effects~${\bf b}_i$,
and~$f({\bf b}_i)$ is the pdf of the random effects~${\bf b}_i$. The
distributions are showed in the equation. There
is no closed-form result for this integral, but it can be numerically
calculated.

\section{Simulation Study}
\label{sec:simu}

The proposed methods were validated with an extensive simulation study which
covers the most commonly seen scenarios for both independent data and clustered
data. For the mean pattern in the simulation setting, only the parametric part was used
instead of the spline bases of time, since the fitting process will be
the same for each candidate value of degree of freedom of the spline
bases, and the same model selection method can be used for selecting
the best value.

\subsection{Independent data}

This study mimics a scenario where each subject has one measure; see the fetal
pancreas length application in Section~\ref{sec:pancreas}.
The data generating model was
\begin{equation*}
y_i = \beta_0 + \beta_1x_{1i} + \beta_2x_{2i} + \varepsilon_i, 
~~i=1,\ldots,n,
\end{equation*}
where~$y_i$ is $i$th response, $x_{1i}$ is a 
$\mbox{Bernoulli}(0.5)$ variable, $x_{2i}$ is a
$\mbox{Uniform}(0, 2)$ variable, 
$(\beta_0, \beta_1, \beta_2)=(1, 1, 1)$, and
$\varepsilon_i$ is a zero-mean normally distributed error term with
heteroscedasticity. Specifically, the standard deviation of $\varepsilon_i$
is~$g(\mu_i)$, where
$\mu_i = \beta_0 + \beta_1x_{1i} + \beta_2x_{2i}$.
Three functional forms were considered for $g$:
$g_1(\mu) = 0.25(\mu - 0.9)$,
$g_2(\mu) = 0.02(\mu^3+1.2)$, and
$g_3(\mu) = 0.1(5\Phi(\mu-2)/0.3)+1)$,
where~$\Phi(\cdot)$ is the cumulative distribution function of the standard
normal distribution.
The specific parameter values in these functions were chosen such
that the function values were positive, the
ratio of the maximum value over the minimum value was around~30, and the
resulting signal-to-noise ratio in the linear regression model was
around~3. Two sample sizes were considered $n \in \{100, 200\}$.

For each configuration, 1000 datasets
were generated. For comparison, both the naive linear regression model with
heteroscedasticity ignored and the proposed GCA with shape-restricted
heteroscedasticity were fitted to each dataset. For the naive model, the
variances of the regression coefficients were obtained using the robust
estimator to account for the
heteroskedasticity~\citep{White:heter0:1980, MacKinnon:some:1985} as
implemented in the R package
\texttt{sandwich}~\citep{Zeileis:various:2020, Zeileis:eco:2004,
  Zeileis:obj:2006}.
In the proposed method, I-spline bases were used to enforce monotonicity.
As more degrees of freedom were needed for fitting more complex patterns
with satisfying accuracy, we picked quadratic I-spline basis with~2, 3,
and~7 degrees of freedom when fitting the three $g(\cdot)$ patterns,
respectively, with evenly spaced internal knots.
Parametric bootstrap was used to calculate the standard
deviation of the estimates and to construct 95\% confidence interval of the
regression coefficients and the error variance curve. The number of
bootstrapping replicates was 1000.

\begin{table}[tbp]
  \centering
  \caption{Summary of simulation results for the independent data scenario:
    SE is the empirical standard error;
    $\hat{\mathrm{SE}}$ is the average of bootstrap standard errors,
    and CP is the empirical coverage percentage of $95\%$
    confidence intervals.}
  \begin{tabular}{cccrrr p{3em} rrrc}
\toprule 
    & & & \multicolumn{4}{c}{naive method ($\times 10^{-2}$)}
    & \multicolumn{4}{c}{proposed method ($\times 10^{-2}$)} \\
    \cmidrule(lr){4-7} \cmidrule(lr){8-11}
$n$ & $g$ & coef & bias & se &
          $\hat{\mathrm{se}}$
          & CP & Bias & SE & $\hat{\mathrm{SE}}$ & CP \\
\midrule 
    100 & $g_1$ & $\beta_0$ & $\phantom{-}$0.6 &  7.5 & 7.4 & 96.0 & $\phantom{-}$0.0 & 2.3 & 2.7 & 97.0 \\
        &       & $\beta_1$ & $-0.4$ &  8.8 &  8.8 & 94.2 & $-0.1$ & 7.1 & 7.0 & 94.5 \\
        &       & $\beta_2$ & $-0.4$ &  8.0 &  7.9 & 93.7 & $-0.0$ & 5.0 & 5.0 & 94.0 \\
     & $g_2$ & $\beta_0$ & $-0.0$ & 10.0 & 9.7 & 93.6 & $-0.1$ & 2.3 & 2.0 & 90.6 \\
        &       & $\beta_1$ & $\phantom{-}$0.1 & 10.8 & 10.5 & 94.8 & $\phantom{-}$0.1 & 6.2 & 6.0 & 94.1 \\
        &       & $\beta_2$ & $\phantom{-}$0.1 & 10.6 &  10.2 & 92.5 & $\phantom{-}$0.2 & 4.3 & 4.1 & 93.5 \\
     & $g_3$ & $\beta_0$ & $-0.1$ &  8.0 & 7.5 & 94.6 & $-0.3$ & 3.7 & 3.4 & 90.0 \\
        &       & $\beta_1$ & $-0.1$ &  9.9 & 9.9 & 94.7 & $-0.1$ & 9.3 & 9.0 & 93.7 \\
        &       & $\beta_2$ & $\phantom{-}$0.2 &  9.1 &  8.4 & 91.9 & $\phantom{-}$0.4 & 6.7 & 6.4 & 93.6 \\ [1ex]
    200 & $g_1$ & $\beta_0$ & $-0.0$ &  5.5 &  5.2 & 93.5 & $-0.0$ & 1.6 & 1.9 & 97.7 \\
        &       & $\beta_1$ & $\phantom{-}$0.0 &  6.2 &  6.2 & 95.0 & $-0.0$ & 5.1 & 4.9 & 94.1 \\
        &       & $\beta_2$ & $-0.0$ &  6.0 &  5.6 & 92.7 & $-0.0$ & 3.5 & 3.5 & 94.6 \\
     & $g_2$ & $\beta_0$ & $\phantom{-}$0.1 &  6.9 &  7.0 & 95.9 & $-0.0$ & 1.5 & 1.4 & 92.3 \\
        &       & $\beta_1$ & $-0.0$ &  7.2 &  7.5 & 95.0 & $\phantom{-}$0.1 & 4.1 & 4.2 & 95.3 \\
        &       & $\beta_2$ & $-0.1$ &  7.4 &  7.3 & 94.5 & $\phantom{-}$0.0 & 2.9 & 2.9 & 94.7 \\
     & $g_3$ & $\beta_0$ & $-0.1$ &  5.5 &  5.3 & 94.8 & $-0.0$ & 2.7 & 2.4 & 92.3 \\
        &       & $\beta_1$ & $\phantom{-}$0.2 &  6.9 &  7.0 & 95.0 & $-0.1$ & 6.3 & 6.2 & 94.5 \\
        &       & $\beta_2$ & $\phantom{-}$0.1 &  6.1 &  5.9 & 94.3 & $\phantom{-}$0.0 & 4.7 & 4.5 & 92.7 \\
\bottomrule
\end{tabular}
\label{table:simu_ind_data}
\end{table}

Table~\ref{table:simu_ind_data} summarizes the empirical bias, empirical
standard error (se), estimated standard error ($\hat{\mathrm{se}}$), and the
coverage percentage (CP) of the 95\% confidence intervals of parameter
estimates. The bias from the proposed method is close to zero under all the
scenarios. The estimated standard deviations from parametric bootstrap
$\hat{\mathrm{se}}$ is close to the empirical value from the proposed
method, and the CP is close to the nominal level 95\%. The point estimate of
$\beta_2$ has lower variation than that for $\beta_1$, which is expected
because the continuous covariate~$x_2$ provides more information than the
binary~$x_1$. As sample size increases, all standard errors decreases.
In comparison with proposed method, the naive method leads to much higher
standard deviations in the regression coefficient estimation. Although the
95\% confidence intervals from the naive method seem to have appropriate
coverage percentage, they are much longer than those from the proposed method
as evident from the standard errors.

\begin{figure}[tbp]
\centering
\includegraphics[width = \textwidth]{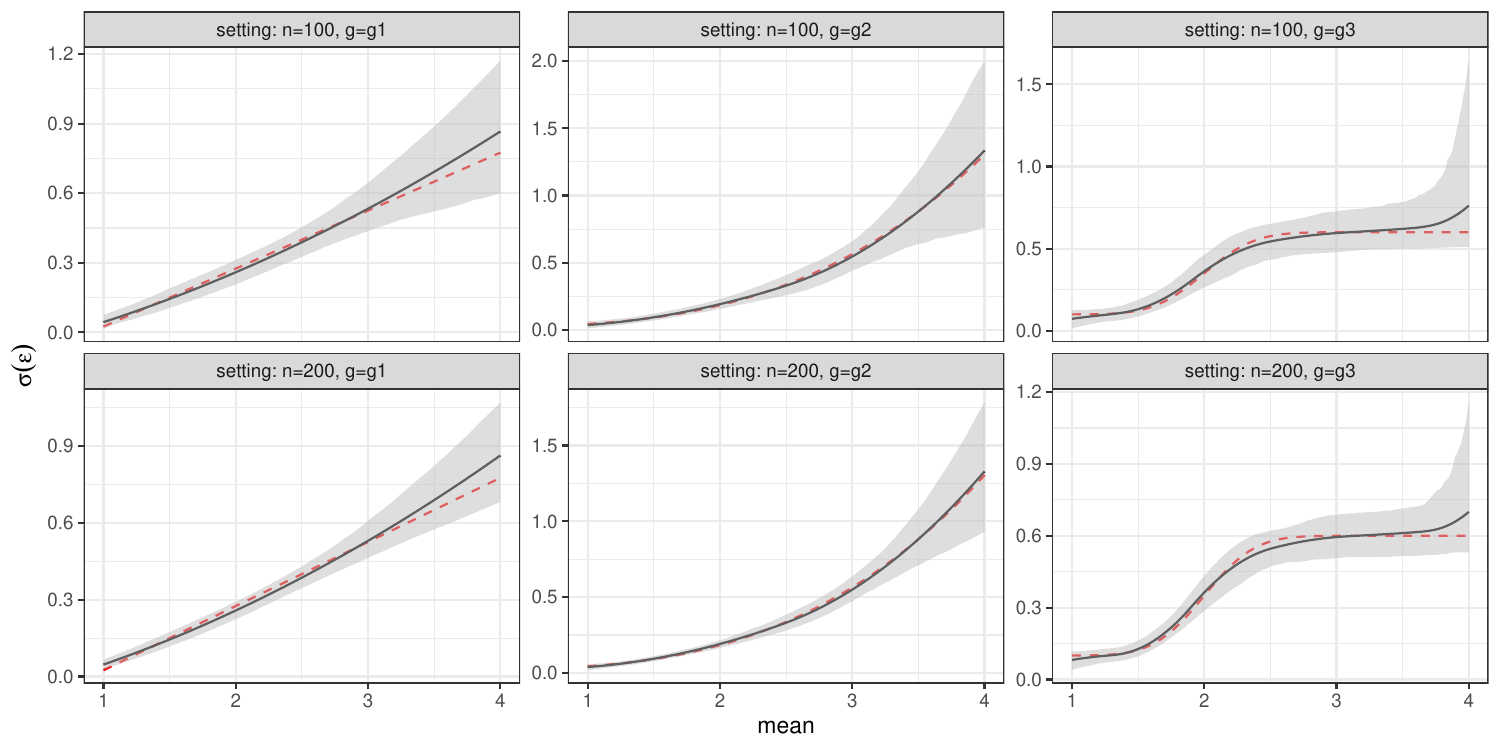}
\caption{Error variance estimation for independent data simulation. The
red dashed line  is the truth, and the black solid line is the estimation. The
grey region is the~95$\%$ point-wise confidence interval.}
\label{fig:ErrVarLm}
\end{figure}

Figure~\ref{fig:ErrVarLm} displays the estimated heteroscedasticity form and
the averaged point-wise 95\% confidence intervals from the proposed method using
parametric bootstrap. The averaged estimates represented by the solid lines are
close to the true curves in dashed lines, and the true
curves lie within the 95\% point-wise confidence intervals.
The wider intervals near the right boundary, especially in the case of $g_3$,
reflect that this is a challenging situation; the functional form is convex
first and concave later, which needs more degrees of freedom to
capture.

\subsection{Clustered data}

For clustered data, we considered a setting where each subject has repeated
measures; see the chicken weight application in Section~\ref{sec:chicken}.
The data generating model was a mixed-effects model with a random effect at the
subject level, 
\begin{equation*}
  y_{ij} = \beta_0+\beta_1x_{1i}+\beta_2x_{2ij}+b_i+\varepsilon_{ij},
  \quad i = 1, \ldots, n, \quad j = 1, \ldots, 5,
\end{equation*}
where $y_{ij}$ is the response variable for the~$j$th observation from the
$i$th subject, $x_{1i}$ is a subject-level covariate generated from
$\mbox{Bernoulli}(0.5)$, $x_{2ij}$ is an observation-level continuous covariate
generated from $\mbox{Uniform}(0, 5)$,
$(\beta_0, \beta_1, \beta_2)=(1, 1, 1)$, $b_i$ is the subject-level random
effect generated from $\mathrm{N}(0, \sigma_{b}^2)$ with $\sigma_{b} = 0.1$,
and~$\varepsilon_{ij}$ is a zero-mean normally distributed error term with
heteroscedasticity. The standard deviation of $\varepsilon_{ij}$ was set to be
$g(\nu_{ij})$, where $g \in \{g_1, g_2, g_3\}$ is same as in the independent data
setting,  $\nu_{ij}$ is either the conditional mean or the marginal mean of the
$j$th observation of the $i$th subject defined in Equation~\eqref{eq:cmean}
and~\eqref{eq:mmean}, respectively. The total number of subjects was set to be
again~$n \in \{100, 200\}$.

Three different estimation methods were compared. The first is the naive method
that fits a linear mixed-effects model with constant error
variance. To better capture the standard error, the robust sandwich
estimator is used~\citep{Ting:Merkle:merDeriv:2018}. The
second method uses the function \texttt{lme()} from the \textsf{R}
package \textbf{nlme}, which provides some preset forms of heteroscedasticity.
For settings where $g_1$ and $g_2$ are used and $\mu$ is the
conditional mean, we used the true setup to specify the error variance
when fitting the model. For other settings, the correct setup is not
available in the \texttt{lme()} function, but we still used a power function
of the conditional mean to specify the heteroscedasticity.
The third method is the proposed method, where quadratic I-spline bases were
used with degrees of freedom~5 for~$g_1$ and $g_2$, and~7
for~$g_3$. Internal knots of spline bases were chosen to be evenly
spaced. The number of replications for parametric bootstrap was 1000. For each
configuration, results were obtained for 1000 datasets.

\begin{table}[tbp]
\centering
\caption{Summary of simulation results when error
  variance changes with the marginal mean: SE is the
  empirical standard error; `$\hat{\mathrm{SE}}$ is the average of bootstrap
  standard errors, and CP is the empirical coverage percentage of 95$\%$
  confidence intervals.}
\begin{tabular}{ccc rrrp{3em}rrrp{3em}rrrc}
\toprule 
  &  &  &  \multicolumn{4}{c}{naive method ($\times 10^{-2}$)} & \multicolumn{4}{c}{lme ($\times 10^{-2}$)}& \multicolumn{4}{c}{proposed method ($\times 10^{-2}$)}\\
\cmidrule(lr){4-7} \cmidrule(lr){8-11} \cmidrule(lr){12-15}
$n$  & $g$ & coef & Bias & SE & $\hat{\mathrm{SE}}$ 
& CP & Bias & SE
& $\hat{\mathrm{SE}}$ & CP & Bias & SE & $\hat{\mathrm{SE}}$ & CP \\
\midrule 
100 & $g_1$ & $\beta_0$ & $\phantom{-}$0.1 & 8.8 & 8.4 & 94.0 & $-1.6$ & 6.9 & 6.6 & 93.4 & $-0.1$ & 6.9 & 6.6 & 93.4 \\ 
   &  & $\beta_1$ & $-0.2$ & 10.0 & 9.5 & 94.2 & $\phantom{-}$0.1 & 8.3 & 8.4 & 95.4 & $-0.2$ & 8.3 & 8.2 & 94.3\\ 
   &  & $\beta_2$ & $-0.1$ & 3.5 & 3.3 & 93.3 & $\phantom{-}$0.2 & 3.0 & 2.8 & 92.3 & $-0.0$ & 3.0 & 2.9 & 92.8\\
  &  & $\sigma_{b}$ & $-0.7$ & 9.9  &  &  &
                                                         $\phantom{-}$0.4
                  & 8.7 &  &  & $-0.9$ & 8.7 &  & \\
  & $g_2$ & $\beta_0$ & $-0.4$           & 9.5  & 9.7 & 95.2 & $-2.9$ & 4.3 & 3.5 & 81.0 & $-0.1$ & 4.0 & 3.7 & 93.0\\ 
      &       & $\beta_1$ & $\phantom{-}$0.1 & 10.4 & 10.4 & 94.6 & $\phantom{-}$0.2 & 5.0 & 5.6 & 96.2 & $-0.1$ & 4.9 & 4.9 & 94.7\\ 
      &       & $\beta_2$ & $\phantom{-}$0.2 & 4.4  & 4.3  & 93.6 &
                                                                    $\phantom{-}$1.1 & 2.6 & 2.2 & 87.6 & $\phantom{-}$0.1 & 2.4 & 2.3 & 93.6\\
  &      & $\sigma_{b}$ & $-0.4$ & 10.8& & &$\phantom{-}$4.6 &2.8 & & & $-0.1$& 5.4 && \\
  & $g_3$ & $\beta_0$ & $-0.1$           & 7.1 & 7.1 & 95.0 & $-3.8$ & 6.4 & 5.1 & 82.5 & $-0.1$ & 5.4 & 5.2 & 93.8\\ 
      &       & $\beta_1$ & $\phantom{-}$0.3 & 9.0 & 9.0  & 95.4 & $\phantom{-}$1.3 & 6.3 & 7.2 & 97.6  & $\phantom{-}$0.1 & 5.8 & 6.1 & 95.2\\ 
      &       & $\beta_2$ & $\phantom{-}$0.0 & 3.1 & 3.1  & 94.5 &
                                                                   $\phantom{-}$1.1
                  & 2.8 & 2.4 & 88.8 & $\phantom{-}$0.0 & 2.5 & 2.6 &
                                                                      95.1\\
  &      & $\sigma_{b}$ & $-1.2$  & 9.2& & &$\phantom{-}$6.0 &6.6 & & &$-0.8$ & 6.7 & & \\ [1ex]
  200 & $g_1$ & $\beta_0$ & $\phantom{-}$0.2 & 6.1 & 6.0 & 94.9 & $-1.3$ & 4.9 & 4.6 & 92.9 & $\phantom{-}$0.2 & 4.8 & 4.7 & 94.0\\ 
      &       & $\beta_1$ & $-0.4$           & 6.8 & 6.8 & 95.3 & $-0.0$ & 5.8 & 5.9 & 95.5 & $-0.3$ & 5.8 & 5.8 & 95.2\\ 
      &       & $\beta_2$ & $\phantom{-}$0.0 & 2.3 & 2.4 & 95.7 &
                                                                  $\phantom{-}$0.3 & 2.0 & 2.0 & 94.3 & $-0.0$ & 2.0 & 2.0 & 94.8\\
      &       & $\sigma_{b}$ & $-0.8$ &8.6 & & &$\phantom{-}0.5$ &7.3 & & &$-0.5$ &7.5 & & \\
   & $g_2$ & $\beta_0$ & $\phantom{-}$0.0 & 6.9 & 6.9 & 94.7 & $-2.7$ & 3.1 & 2.5 & 75.7 & $\phantom{-}$0.1 & 2.8 & 2.7 & 93.7\\ 
      &       & $\beta_1$ & $-0.0$           & 7.3 & 7.4 & 94.8 & $\phantom{-}$0.2 & 3.7 & 4.0 & 96.3 & $-0.1$ & 3.6 & 3.5 & 94.2\\ 
      &       & $\beta_2$ & $\phantom{-}$0.0 & 3.1 & 3.1 & 94.3 &
                                                                  $\phantom{-}$1.0 & 1.8 & 1.6 & 87.8 & $-0.0$ & 1.6 & 1.6 & 94.1\\
  &      & $\sigma_{b}$ & $-0.9$& 9.1& & &$\phantom{-}$4.9 &2.0 & & &$\phantom{-}$0.8 & 3.9 & & \\
   & $g_3$ & $\beta_0$ & $\phantom{-}$0.1 & 4.9 & 5.0 & 95.2 & $-3.9$ & 4.6 & 3.6 & 74.6  & $\phantom{-}$0.0 & 3.8 & 3.7 & 94.5\\ 
       &       & $\beta_1$ & $-0.0$           & 6.3 & 6.4 & 95.2 & $\phantom{-}$1.3 & 4.8 & 5.1 & 96.0 & $\phantom{-}$0.1 & 4.5 & 4.3 & 94.0\\ 
       &       & $\beta_2$ & $-0.0$           & 2.2 & 2.2 & 95.4 &
                                                                   $\phantom{-}$1.1 & 2.0 & 1.7 & 86.4 & $-0.0$ & 1.8 & 1.8 & 95.5\\
   &      & $\sigma_{b}$ &$-1.3$  &8.1 & & &$\phantom{-}7.1$ &4.8 & & & $-0.5$& 5.7 & & \\
\bottomrule
\end{tabular}
\label{tab:simu_clustered1}
\end{table}

\begin{table}[tbp]
\centering
\caption{Summary of simulation results when the error
  variance changes with the conditional mean: SE is empirical standard error;
  $\hat{\mathrm{SE}}$ is the average of bootstrap standard errors;
  and CP is the empirical coverage percentage of 95\% confidence interval.}
\label{tab:simu_clustered2}
\begin{tabular}{ccc rrrp{3em}rrrp{3em}rrrc}
\toprule 
  &  &  &  \multicolumn{4}{c}{naive method ($\times 10^{-2}$)} & \multicolumn{4}{c}{lme ($\times 10^{-2}$)}& \multicolumn{4}{c}{proposed method ($\times 10^{-2}$)}\\
\cmidrule(lr){4-7} \cmidrule(lr){8-11} \cmidrule(lr){12-15}
$n$  & $g$ & coef & Bias & SE & $\hat{\mathrm{SE}}$ 
& CP & Bias & SE
& $\hat{\mathrm{SE}}$ & CP & Bias & SE & $\hat{\mathrm{SE}}$ & CP \\
\midrule 
100 & $g_1$ & $\beta_0$ & $-0.3$           & 8.5 & 8.5 & 95.1 & $-1.8$ & 6.7 & 6.6 & 93.9 & $-1.3$ & 6.8 & 6.8 & 94.4  \\ 
    &       & $\beta_1$ & $\phantom{-}$0.5 & 9.4 & 9.5  & 94.6 & $\phantom{-}$0.7 & 8.0 & 8.3 & 95.3 & $\phantom{-}$0.5 & 8.0 & 8.3 & 95.7 \\ 
    &       & $\beta_2$ & $\phantom{-}$0.1 & 3.4 & 3.3  & 95.1 &
                                                                 $\phantom{-}$0.4 & 2.9 & 2.8 & 94.9 & $\phantom{-}$0.3 & 2.9 & 2.9 & 94.5 \\
   &      & $\sigma_{b}$ &$-1.5$  &9.4 & & &$-0.3$ &8.4 & & &$-1.4$ &8.7  & & \\
   & $g_2$ & $\beta_0$ & $\phantom{-}$0.4 & 9.8  & 9.7 & 94.2 & $-2.5$ & 4.3 & 3.5 & 83.1 & $-0.5$ & 4.1 & 3.8 & 92.7  \\ 
      &       & $\beta_1$ & $-0.2$           & 10.3 & 10.4 & 94.4 & $\phantom{-}$0.1 & 5.1 & 5.6 & 97.2 & $-0.4$ & 5.0 & 5.0 & 94.8 \\ 
      &       & $\beta_2$ & $-0.1$           & 4.4  & 4.3  & 93.9 &
                                                                    $\phantom{-}$0.9 & 2.6 & 2.2 & 87.5 & $\phantom{-}$0.0 & 2.4 & 2.3 & 92.7 \\
   &      & $\sigma_{b}$ & $-0.7$ &10.7 & & &$\phantom{-}4.5$ &3.0 & & &$-0.1$ &5.4  & & \\
   & $g_3$ & $\beta_0$ & $\phantom{-}$0.0 & 7.4 & 7.1 & 93.7 & $-3.7$ & 6.6 & 5.1 & 81.2 & $-0.3$ & 5.6 & 5.3 & 94.0  \\ 
      &       & $\beta_1$ & $-0.3$           & 9.4 & 9.0  & 93.9 & $\phantom{-}$0.9 & 6.4 & 7.2 & 97.6 & $-0.5$ & 6.2 & 6.1 & 94.6  \\ 
      &       & $\beta_2$ & $\phantom{-}$0.2 & 3.2 & 3.1  & 94.0 &
                                                                   $\phantom{-}$1.2 & 3.0 & 2.4 & 87.3 & $\phantom{-}$0.0 & 2.7 & 2.6 & 95.4  \\
   &      & $\sigma_{b}$ & $-1.9$ & 9.0& & &$\phantom{-}6.0$ &6.7 & & &$-1.0$ &6.7  & & \\ [1ex]
  200 & $g_1$ & $\beta_0$ & $\phantom{-}$0.0 & 6.0 & 6.0 & 94.9 & $-1.4$ & 4.9 & 4.6 & 92.4 & $-1.0$ & 4.8 & 4.8 & 94.7 \\ 
      &       & $\beta_1$ & $\phantom{-}$0.0 & 6.7 & 6.7 & 94.7 & $\phantom{-}$0.4 & 5.8 & 5.9 & 95.4 & $\phantom{-}$0.3 & 5.8 & 5.8 & 94.2  \\ 
      &       & $\beta_2$ & $\phantom{-}$0.0 & 2.4 & 2.4 & 95.1 &
                                                                  $\phantom{-}$0.3 & 2.0 & 2.0 & 94.3 & $\phantom{-}$0.2 & 2.0 & 2.1 & 95.6 \\
   &      & $\sigma_{b}$ &$-1.9$  &8.1 & & &$\phantom{-}0.1$ &7.2 & & &$-0.9$ & 7.6 & & \\
   & $g_2$ & $\beta_0$ & $\phantom{-}$0.1 & 6.9 & 6.9 & 93.8 & $-2.9$ & 3.0 & 2.5 & 73.5 & $-0.7$ & 2.8 & 2.7 & 92.7  \\ 
      &       & $\beta_1$ & $-0.1$           & 7.4 & 7.4 & 95.7 & $\phantom{-}$0.4 & 3.7 & 4.0 & 96.8 & $-0.2$ & 3.6 & 3.5 & 95.2  \\ 
      &       & $\beta_2$ & $-0.0$           & 3.1 & 3.1 & 94.5 &
                                                                  $\phantom{-}$1.1 & 1.8 & 1.6 & 85.0 & $\phantom{-}$0.0 & 1.7 & 1.6 & 94.9 \\
   &      & $\sigma_{b}$ &$-0.9$  & 9.4& & &$\phantom{-}4.6$ & 2.0& & &$\phantom{-}0.3$ & 4.1 & &\\
   & $g_3$ & $\beta_0$ & $\phantom{-}$0.1 & 5.3 & 5.0 & 93.4 & $-4.1$ & 4.9 & 3.6 & 72.8 & $-0.3$ & 3.9 & 3.8 & 94.0  \\ 
       &       & $\beta_1$ & $-0.2$           & 6.8 & 6.4 & 93.6 & $\phantom{-}$1.3 & 4.8 & 5.1 & 95.4 & $-0.1$ & 4.4 & 4.3 & 94.7  \\ 
       &       & $\beta_2$ & $\phantom{-}$0.0 & 2.4 & 2.2 & 93.4 &
                                                                   $\phantom{-}$1.2 & 2.1 & 1.7 & 83.8 & $-0.1$ & 1.9 & 1.8 & 94.3  \\
   &      & $\sigma_{b}$ &$-1.5$  &8.1 & & &$\phantom{-}$7.2 &5.1 & & & $-0.4$&5.6  & & \\
\bottomrule
\end{tabular}
\end{table}

Tables~\ref{tab:simu_clustered1} and~\ref{tab:simu_clustered2}
summarize the simulation results for the variance as a function of marginal mean
and conditional mean, respectively. All three
methods seem to give unbiased point estimates for the regression coefficients,
as they all have correct specification of the regression model. Their
differences are in their uncertainty levels and coverage percentages of the
95\%
confidence intervals. The naive method has the worst
performance since it did not consider the heteroscedasticity at all. The method
with function \texttt{lme()} performs better than the naive method, but is still
not
satisfactory due to the misspecified heteroscedasticity form through the
limited
choices offered by \texttt{lme()}. Even in settings where the
heteroscedasticity
is correctly specified by the power function~$g_2()$ with conditional mean as
the index variable, see Table~\ref{tab:simu_clustered2}, its coverage
percentages for $\beta_2$ are much lower than 95\%. In contrast, in all three
settings, the proposed method gives point estimates with lower standard errors
as well as confidence intervals with coverage percentage close to 95\%.
As the sample size~$n$ increases, the performance becomes better as
expected.

Also reported in Tables~\ref{tab:simu_clustered1} and~\ref{tab:simu_clustered2}
are estimates of the standard deviation $\sigma_b$ of the random effects.
It is known that the standard errors of the random-effects variance parameters
are hard to get and confidence intervals constructed from profile likelihood
or parametric bootstrap are preferred~\citep{Bolker:wald:2016}.
So here we focus on the point estimate of~$\sigma_b$.
When the heteroscedasticity takes more complicated forms such as $g_2$ and
$g_3$, the proposed method has much smaller bias than the \texttt{lme()} method.
In some settings such as $g = g_2$; the \texttt{lme()} estimates of $\sigma_b$
have bias but lower variation compared to those from the proposed method.
This echos that caution is needed when using standard errors of the random
effect variance. The empirical standard errors of the point estimates from the
proposed method decrease as the sample size increases, but apparently not at
the
rate of $1/\sqrt{n}$, suggesting that a larger sample size is needed for the
asymptotic properties of the random-effect variance estimator to hold.

\begin{figure}[tbp]
\includegraphics[width = \textwidth]{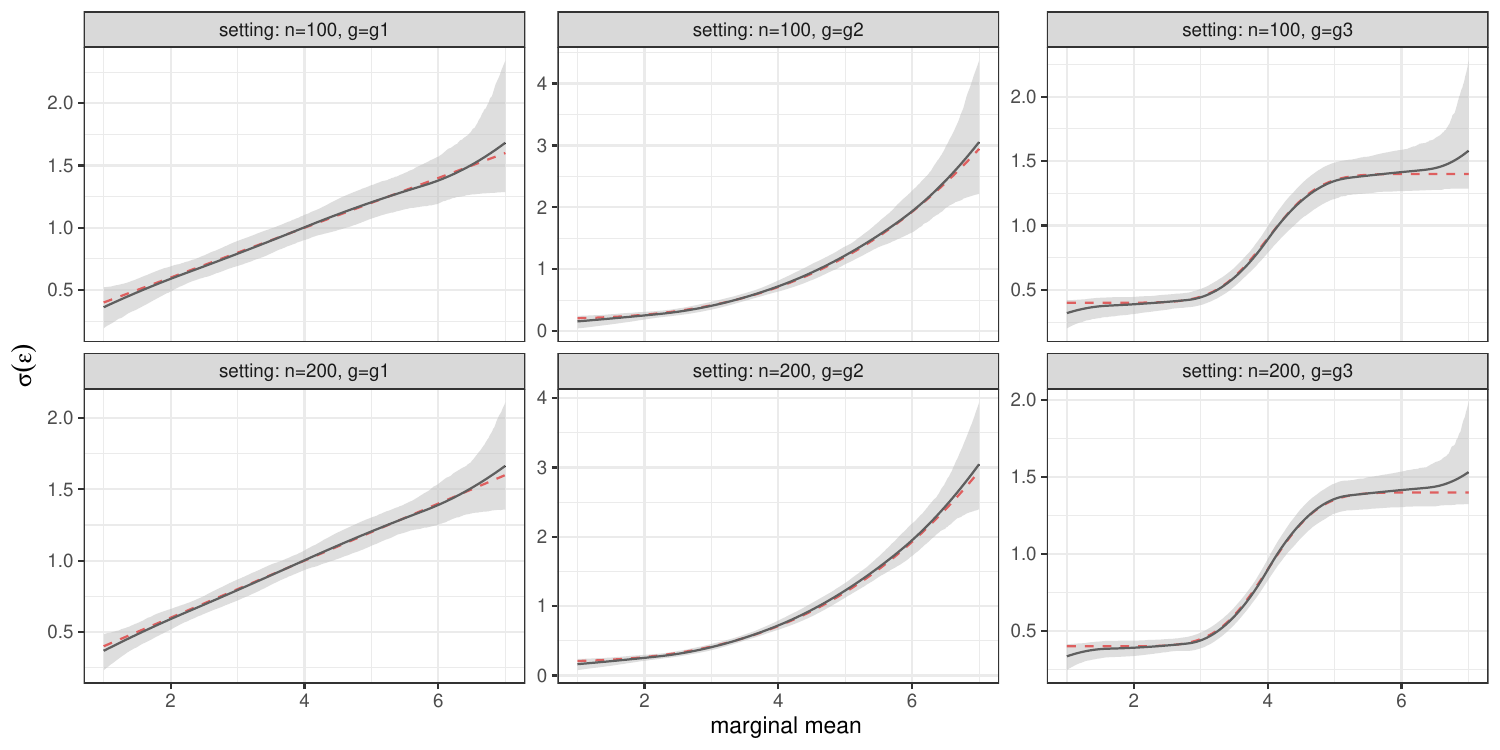}
\includegraphics[width = \textwidth]{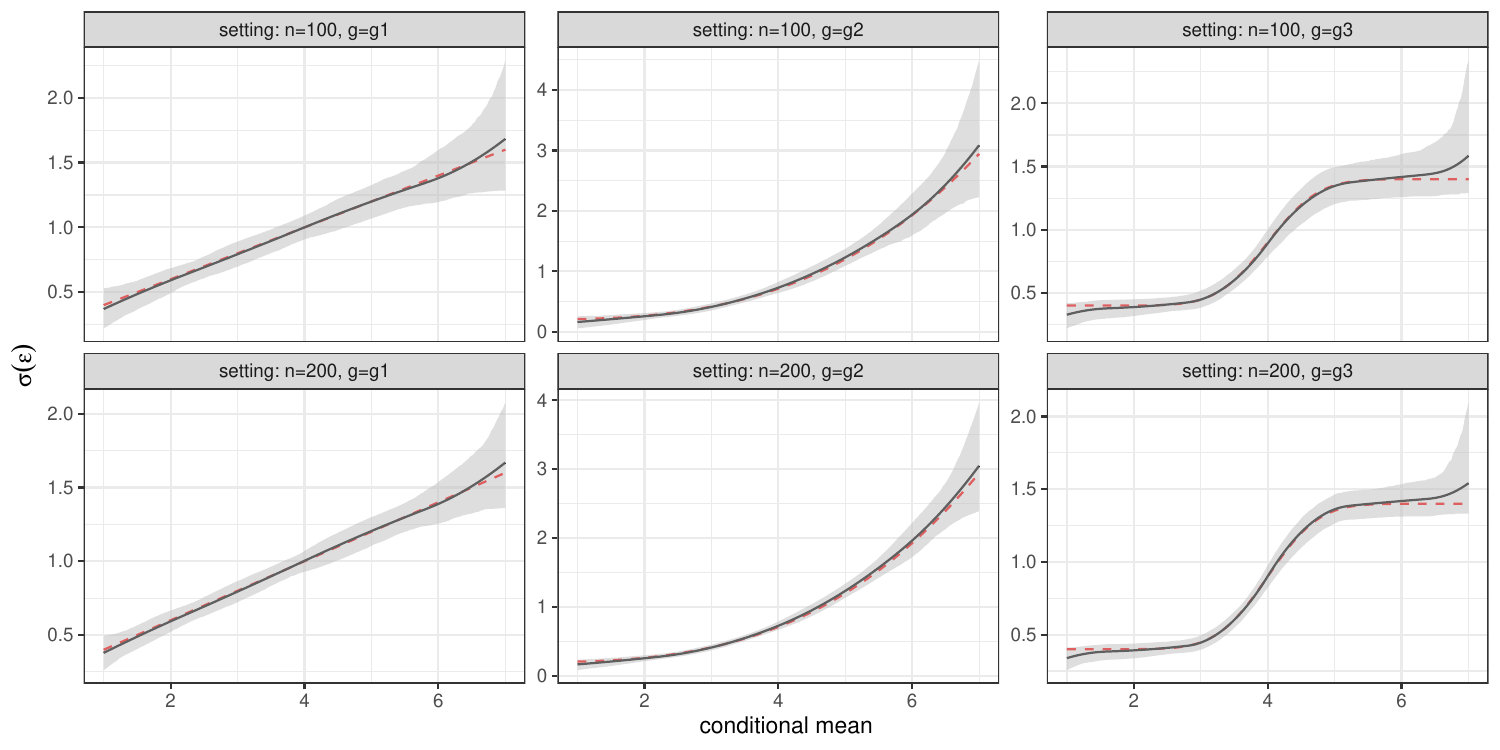}
\caption{Heteroscedasticity estimation for clustered data when it changes with
  marginal mean (upper) and conditional mean (lower).
  The red dashed line is the truth, and the black solid line is  
  the estimation. The grey region is the averaged~95$\%$
  point-wise confidence interval.}
\label{fig:ErrVarLme}
\end{figure}

Figure~\ref{fig:ErrVarLme} displays the fitted heteroscedasticity from the
proposed method. The two panels show the results with the index variables being
the marginal mean and the conditional mean, respectively.
Similar to the independent data scenarios, the estimated curve is close to 
the true curve, and is within the averaged~$95\%$ point-wise confidence
intervals. The intervals are narrower than those in the independent data
scenarios as the repeated measures provide more information.

\section{Applications}
\label{sec:real}

\subsection{Fetal Pancreas Length}
\label{sec:pancreas}

Fetus pancreatic dysplasia and hypertrophy, i.e., abnormality of fetal
pancreas, are associated with congenital
malformations~\citep{quinn2012ontogeny,lyndon1989sonographic}.
The fetal pancreas growth curve is a critical tool for prenatal screening
for  disorders. The growth pattern of fetal pancreas' lengths during the
prenatal period has been investigated~\citep{desdicioglu2010foetal,
elzbieta2005morphometry, hata1988ultrasonographic}, but not the changing
variation of the measurements. The proposed method allows capturing the
fetal pancreas' growing patterns in both the mean level and the variation
level.

\begin{figure}[tbp]
\includegraphics[width = \textwidth]{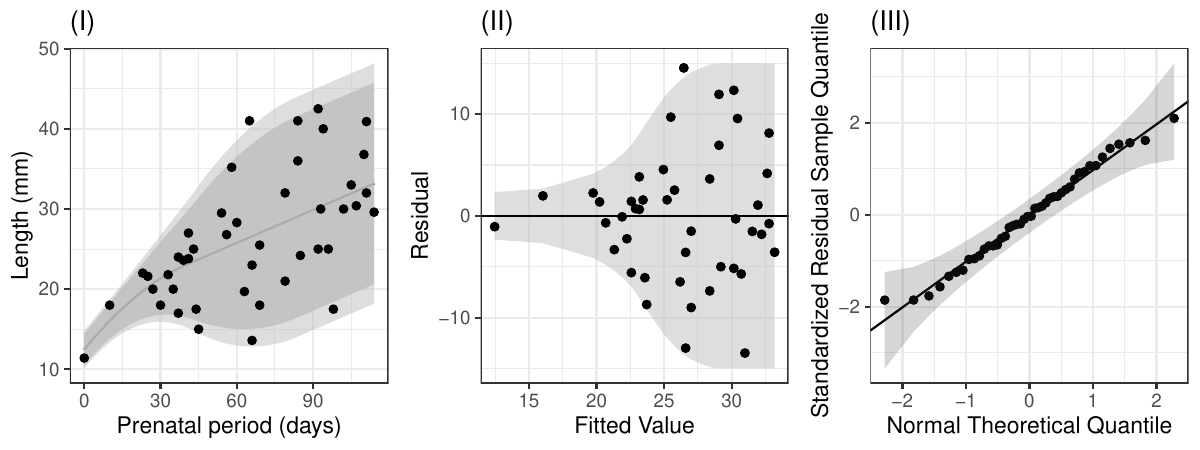}
\centering
\caption{Pancreas length data analysis: (I) original data with estimated mean
    and $90\%$ and $95\%$ point-wise confidence intervals; (II) the residual
    versus fitted mean plot with $95\%$ point-wise confidence interval;
    (III) Q-Q plot of standardized residual.}
\label{fig:pancrea}
\end{figure}

One of the authors of this paper, Dr. Wenling Song from the Second
Hospital of Jilin University, collected the healthy pancrea length
data and provided it for analysis in this paper. The data were
collected from~44 pregnant women at different stages of pregnancy
who visited the Second Hospital of Jilin University in China during April to
July of~2012. The data is provided by Dr.Wenling Song, who is also the
author of this paper. No patient showed any external pathology or
anomaly. The dataset contains a single measure of the fetal pancreas' length from
each patient. Figure~\ref{fig:pancrea} (I) shows the pancreas' lengths in
millimeters versus the pregnant duration in days. It is reasonable
to assume that the growth speed
slows down as the fetus matures, in which case the growth curve would be
increasing and concave. It is also reasonable to assume that the
variation increases with time. Therefore, we use C-splines to model the mean
growth level and use I-splines for the heteroscedasticity. Specifically, the
model is
\begin{align*}
  y_i &= \beta_0+ \beta_1 t_i +
  \sum_{k = 1}^{K_1} \beta_{k+1} C_k(t_i) +  \varepsilon_i,\\
  \varepsilon_i &~ \sim \mathrm{N}\left(0, \quad\left(\theta_0 +
  \sum_{k = 1}^{K_2}\theta_kI_k(t_i)\right)^2\right),
\end{align*}
where $y_i$ is the $i_{th}$ length measurement, $t_i$ is the
corresponding time with linear coefficient $\beta_1$,
$C_k(t_i)$ is the $k$th C-splines basis evaluated at $t_i$ with
coefficient $\beta_{k+1}$, $k = 1, \ldots, K_1$,
$\varepsilon_i$ is the independent error term with heteroscedasticity, and
$I_k(t_i)$ is the $k$th I-spline basis evaluated at $t_i$ with coefficient
$\theta_k$, $k = 1, \ldots, K_2$. The spline bases were selected using
the method stated in Section~\ref{sec:ModelSelection}. The internal
knots of both spline bases were evenly spaced. The degrees of freedom
$K_1$ and $K_2$ for the C-splines and I-splines, respectively, were
both chosen to be 4 by BIC.

Figure~\ref{fig:pancrea} shows fitted results for the growth curve of the
pancreas length. By shape restrictions, the fitted curve is increasing and
concave while the variance is increasing over time. The estimated curve and the
point-wise confidence intervals in panel (I) accurately capture the mean and variation
pattern. The residual plot with fitted~$95\%$ confidence intervals in panel (II)
shows good performance on estimating the heterogeneous pattern of error
standard deviation. The Q-Q plot of the standardized residuals in panel (III) shows no
alarming deviation from the normality.

The fitted results show accurate estimate of the pancrea growth curve
along with pregnancy days, including the mean and the quantiles. This
can provide a better guidance of screening abnormality specifically
for babies in that area. 

\subsection{Chicken Weight}
\label{sec:chicken}

\begin{figure}[tbp]
\includegraphics[width = \textwidth]{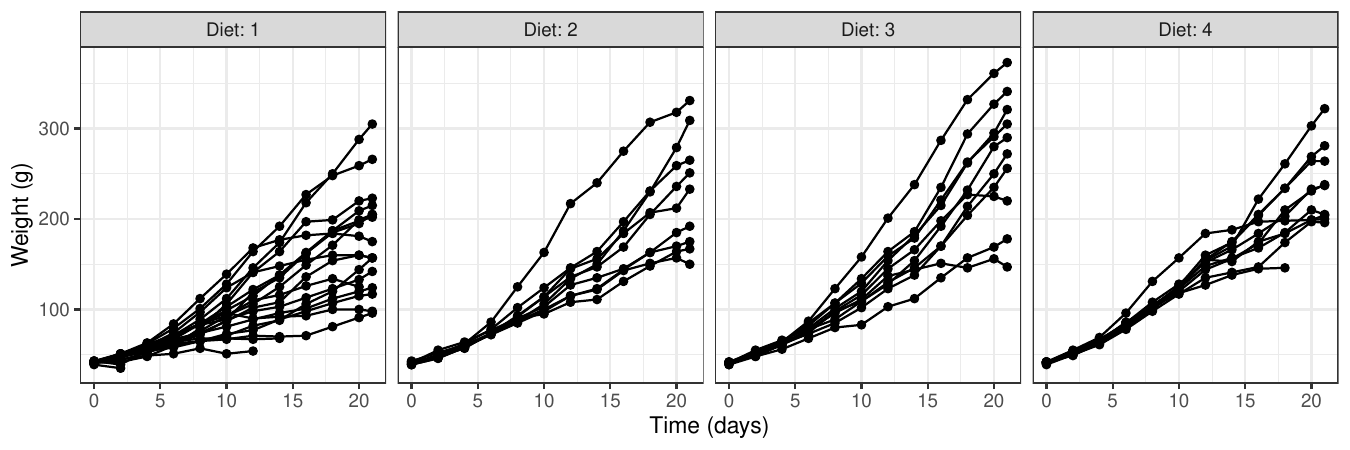}
\centering
\caption{Weights over time in days for chicks in the four diet
groups in the chicken weight dataset.}
\label{fig:chick}
\end{figure}

\begin{figure}[tbp]
  \centering
  \includegraphics[width = \textwidth]{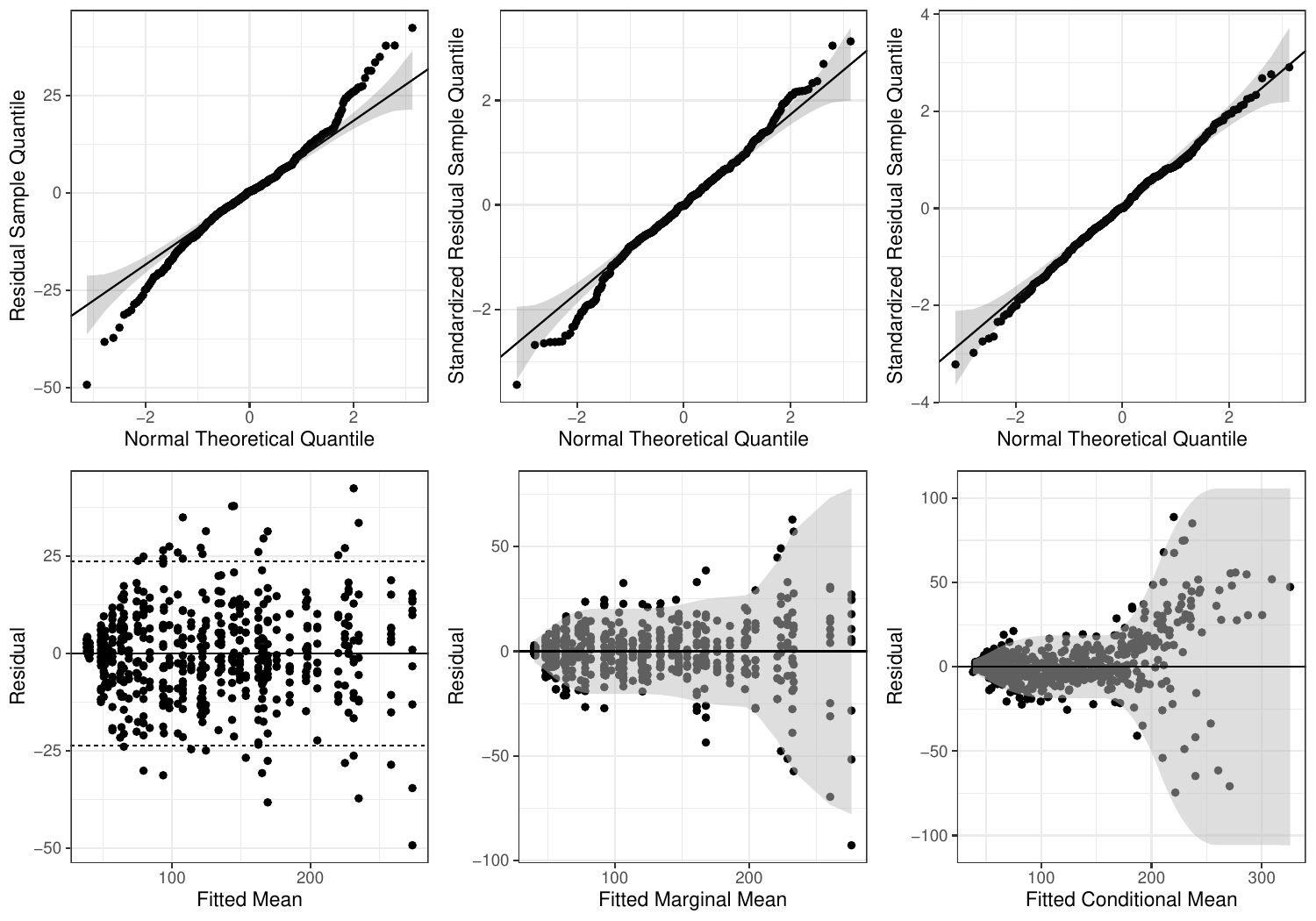}
  \caption{Diagnostics of chicken weight data analysis.
    Left: naive model; Middle: heteroscedastic model with variance
    changing with marginal mean; Right: heteroscedastic model with variance
    changing with conditional mean.  Upper: Q-Q plots of
    standard residual; Lower: residual versus fitted means, with the 95\%
    point-wise confidence intervals.
    }
  \label{fig:Chickfit}
\end{figure}

\begin{figure}[tbp]
  \centering
  \includegraphics[width = \textwidth]{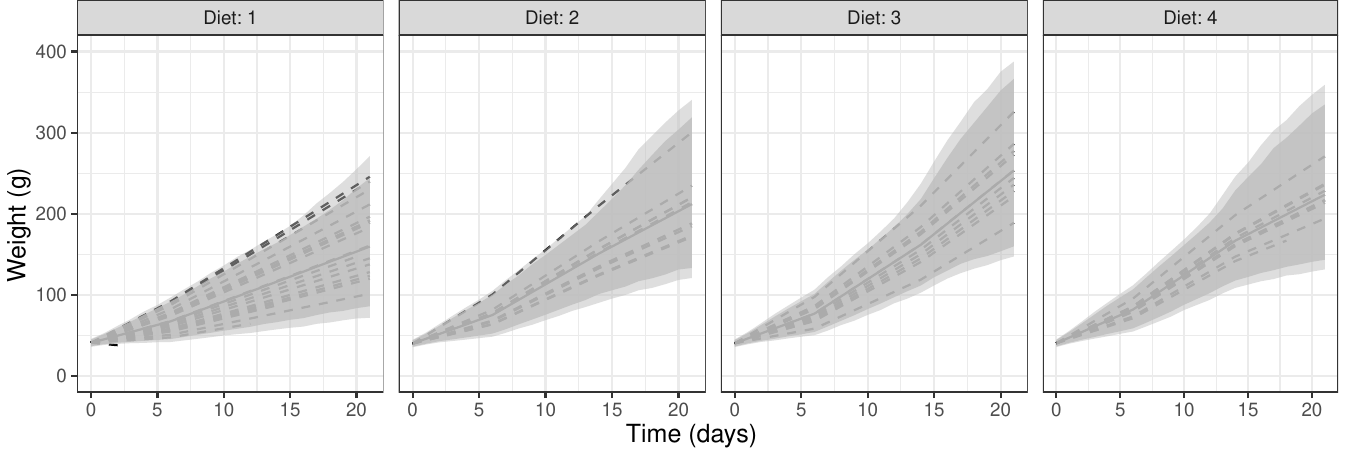}
  \caption{Fitted growing curves for all four Diet groups overlaid with
    the point-wise 90\% and 95\% confidence intervals . The solid lines
    are the fixed effects, and the dashed lines represent the chicks with random
    effects on Time. 
  }
  \label{fig:chick_gc}
\end{figure}

The chicken weight data, which is available in \textsf{R} package
\textbf{datasets}, is a classic example of clustered data for
GCA~\citep[][p.~4]{Hand:practical:1996}. It
contains the body weights in grams of~50 chicks measured at birth date and every
other day thereafter until day~20, plus an additional measurement on
day~21. These~50 chicks were divided into four groups to have different
protein diets, and the scatter plots of weight versus time
for each group are shown in
Figure~\ref{fig:chick}. We removed Chick No. 24 as an outlier from the original
dataset since its weight stopped increasing after day 6. All
four groups show increasing trends on both the mean the variation
level. Previous studies focused on
building regression models on weight gain~\citep{Hand:practical:1996}, i.e., the
weight change, but no work has been done to directly capture the
heteroscedasticity in GCA.

We fitted Model~\eqref{clustered_model} to this dataset. The model includes
fixed effects consisting of linear I-spline bases of time and their
interactions
with the diet, as well as a chick-level random effects on the slope of time.
Heteroscedasticity is characterized by an I-spline of an index variable,
which is either the marginal mean or the conditional mean of the linear effects
model. Specifically, the model is
\begin{align*}
    y_{ij} &= \beta_0 + \sum_{k = 1}^{K_1}\left(\beta_kI_{1k}(t_{ij}) +
      \beta^{(2)}_kI_{1k}(t_{ij})D^{(2)}_{i} +
      \beta^{(3)}_kI_{1k}(t_{ij})D^{(3)}_{i} +
      \beta^{(4)}_kI_{1k}(t_{ij})D^{(4)}_{i}\right) + b_it_{ij} + 
    \varepsilon_{ij},\\
    b_i &~\sim \mathrm{N}(0, \sigma_b^2),\\
    \varepsilon_{ij} &~ \sim \mathrm{N}\left(0, \quad \left(\theta_0
    +\sum_{k = 1}^{K_2} \theta_kI_{2k}(\nu_{ij})\right)^2\right),
\end{align*}
where $y_{ij}$ is the weight of chick $i$ at time $t_{ij}$,
$\{I_{1k}(t_{ij}): k = 1, \ldots, K_1\}$ is a set of I-splines bases with
$K_1$~degrees of freedom used in the mean model,
$(D^{2}_{ij}, D^{(3)}_{ij}, D^{(4)}_{ij})$ are the
dummy variables of diet using diet 1 as the reference level, $b_i$ is
a normally distributed chick-level random effect with variance
$\sigma_b^2$, the error term $\varepsilon_{ij}$ is normal with mean zero and
variance changing with index variable~$\nu_{ij}$ (either marginal or
conditional
mean), $\{I_{2k}(\mu_{ij}): k = 1, \ldots, K_2\}$ is a set of I-spline bases
with $K_2$ degrees of freedom used in the heteroscedasticity model, 
and the regression coefficients to be estimated are
$\{\beta_k, \beta^{(2)}_k,  \beta^{(3)}_{k}, \beta^{(4)}_k: k = 1, \ldots, K_1\}$
and~$\{\theta_k: k = 1, \ldots, K_2\}$. The internal knots of spline
bases were always chosen to be evenly spaced.

We first need to decide whether to use the marginal mean or the conditional
mean
as the index variable in the heteroscedasticity model. For both situations, the
BIC chose the same number of degrees of freedom of the I-splines.
The I-spline bases for the mean pancreas length had
degree~0 with 3~degrees of freedom. The I-spline bases for the variance
model had degree~1 and 9~degrees of freedom. Since both the
marginal mean model and the conditional mean model had the same number of
parameters, we can choose the best model by only comparing the log-likelihood.
The conditional mean model had a log-likelihood of $-2164.899$,
which is significantly higher than that of the marginal mean model,
$-2199.183$. Both of them were much higher than $-2315.220$, the
log-likelihood
of the naive model that did not consider heteroscedasticity.

Figure~\ref{fig:Chickfit} shows the diagnostic plots for the three models.
For the naive model, the residual plot suggests increasing variance as fitted
value increases and the Q-Q plot suggest heavier tail than the normal
distribution. After considering heteroscedasticity with the proposed method,
the model with variance changing with the marginal mean model still has heavy
tail problem as seen from the Q-Q plot. The model with variance changing with
the conditional mean, however, shows no obvious deviation from the normal
distribution in the Q-Q plot of the standardized residuals. These diagnostics
are consistent with the model comparison results in terms of
log-likelihood.

Point-wise quantiles are of important practical value in GCA since they can
be used as reference to check if an individual is in the normal range. For
this linear mixed-effects model, we approximated the quantiles by generating
10,000 individuals using the random effects and for each of them, simulating
their grow curve using the fitted model. The upper and lower 5\% quantiles
will give us 90\% confidence interval, and the upper and lower 2.5\% quantiles
will form the 95\% confidence interval. Finally, the fitted growth curves for
the four diet groups along with their point-wise 90\% and 95\% confidence
intervals from the model with residual variance changing with the
conditional mean are shown in Figure~\ref{fig:chick_gc}. Also overlaid are
the fitted individual curves for all the chicks in each diet group. It indicates
that diet~3 works best for the chicks with fastest growing speed and highest weight
at the end of the experiment, while diet~1 is the least
favorable. The estimated results along with other information such as
diet cost, can help the farmer decide which diet to use to generate
the highest profit. The quantile estimation can also provide guidance
on unhealthy chicken screening, and stop potential disease spead at
early stage.

\section{Discussion}
\label{sec:discussion}

Shape-restricted splines provide great flexibility in incorporating prior
knowledge about the shapes of the curves to be fitted. With the recently
available \textsf{R} package \textbf{splines2}~\citep{wang2021shape}, such
fitting is facilitated in routine data analysis. GCA is an important area where
shape restrictions often need to be enforced. In addition to the mean growth
level, the heteroscedasticity can also have shape restrictions. Such shape
restrictions are enforced through constrained optimizations in an iteratively
reweighted fitting procedure, which takes advantage of existing software
routines that allow weights.
For clustered data, the variance of the error term can be changing with either
the marginal mean or the conditional mean. In the latter case, the likelihood
is hard to calculate as the marginal distribution of the response vector is no
longer within the multivariate normal family. This is not too much of an
inconvenience because the iteratively reweighted fitting procedure
does not need to evaluate this likelihood. It is only needed in
calculating model comparison criteria, which only needs to be done
once for each fitted model.

Although proposed in the context of GCA, our method is applicable to the
general setting of linear mixed-effects models or multi-level models with 
shape restricted heteroscedasticity. In fact, our simulation studies were done
in a general setting. More accurate point and interval estimators are expected
when the heteroscedasticity is appropriately accounted for. The parametric
bootstrap for inferences also works well in providing valid uncertainty
measures
for the estimated parameters. Alternative approaches to shape restrictions are
possible. For example, isotonic regression~\citep{Barlow:Brunk:isotonic:1972,
  Wang:Li:isotonic:2008} can be used to enforce monotonicity, but its
implementation and extension to curvature restrictions may not be as simple.
Beyond linear models, application of shape restrictions in generalized
additive models for location, scale, and shape~\citep{rigby2005generalized} or
quantile regressions~\citep{wei2006quantile} merits further
investigation. More complicated shape restrictions beyond monotonicity
and concavity/convexity can be discussed for future work, especially
when the shape is a combination of concavity and convexity such as the
function $g_3$ in our simulation setting, or the variance is bigger on
both end of the boundaries.

%\begin{appendix}

%\section{Appendix section}

%\end{appendix}

\bibliographystyle{nessart-number}
\bibliography{growth}

%\atColsBreak{\pagediscards}
\end{document}